# Universal method for the synthesis of arbitrary polarization states radiated by a nanoantenna

Francisco J. Rodríguez-Fortuño, Daniel Puerto, Amadeu Griol, Laurent Bellieres, Javier Martí and Alejandro Martínez*

*Nanophotonics Technology Center, Universitat Politècnica de València, Valencia (Spain)*

*Correspondence to: amartinez@ntc.upv.es

**Abstract:**

**Optical nanoantennas efficiently convert confined optical energy into free-space radiation. The polarization of the emitted radiation depends mainly on nanoantenna shape, so it becomes extremely difficult to manipulate it unless the nanostructure is physically altered. Here we demonstrate a simple way to synthetize the polarization of the radiation emitted by a single nanoantenna so that every point on the Poincaré sphere becomes attainable. The nanoantenna consists of a single scatterer created on a dielectric waveguide and fed from its both sides so that the polarization of the emitted optical radiation is controlled by the amplitude and phase of the feeding signals. Our nanoantenna is created on a silicon chip using standard top-down nanofabrication tools, but the method is universal and can be applied to other materials, wavelengths and technologies. This work will open the way towards the synthesis and control of arbitrary polarization states in nano-optics.**

A fundamental property of coherent light is its polarization, which is defined by the orientation of its electric field, and determines the forces which light exerts on charges. Polarization is therefore an essential concept to describe the interaction of light with matter, and as such, its generation and measurement is crucial in polarimetry applications ranging from astronomy [1] to thin film ellipsometry. Polarization also plays a key role in the efficient reception and detection of light [2] as well as in quantum optics [3–5]. Recent work on circularly polarized light, which carries spin angular momentum, has revealed interesting applications including generation of spinning forces on particles [6], detection of molecular spinning [7] and novel methods for ultrafast magnetic storage [8,9], requiring a very fast switching between left and right handed circular polarization.

The emission and/or reception of light can be achieved with the use of optical nanoantennas that efficiently convert confined optical energy into free-space radiation [10–16]. In the far-field, the radiation emitted by a nanoantenna has a given polarization, determined by the radiation angle, wavelength and properties of the specific nanostructure, specially its geometry. The polarization is essentially fixed for a given wavelength unless the nanostructure is physically or mechanically altered. Therefore, achieving a fast and tunable control in the emitted polarization of a nanoantenna has remained elusive so far. In this work we describe a universal method to achieve this. Although the concept of nanoantennas in plasmonics usually refers to structures which convert localized near fields into radiating waves, and vice versa, in this work we consider a traditional definition of antenna as that of a transitional structure between free-space and a guiding device [17]. The idea stems from a recent series of works in which the direction of propagation of surface plasmons is selected by the polarization of the incident light [18–22]. This relationship between polarization and directionality of guided modes is not limited to plasmonics, and can be extended to electromagnetic waves in general [19] including dielectric waveguides. We exploit the relationship by considering the reciprocal scenario, in which guided light incoming from different directions results in far-field radiation with different polarizations. We demonstrate a dual-input nanoantenna that can synthesize any desired polarization state across the whole Poincaré sphere by appropriately tuning the relative amplitude and phase of its two feeding waveguides.

The method is summarized in Fig. 1. Consider a nanoantenna with two input waveguides which, when fed from one of them, is designed to radiate in the normal direction an electric field $\mathbf{E_a}$ linearly polarized at 45º to the symmetry plane of the structure (Fig. 1a). The present discussion is extensible to any other polarization for $\mathbf{E_a}$, as long as it is not parallel to the symmetry plane, but we chose a linear polarization at 45º due to its simplicity and mathematical elegance. The radiated electric field can be written as $\mathbf{E_a} = E_0 \left( \frac{\hat{\mathbf{y}} + \hat{\mathbf{x}}}{\sqrt{2}} \right) = E_0 \hat{\mathbf{a}}$ assuming an $\exp(-i\omega t)$ time dependence, where $\hat{\mathbf{x}}$ is normal to the symmetry plane. Owing to symmetry considerations, when fed by the other waveguide, the emitted radiation will be $\mathbf{E_b} = E_0 \left( \frac{\hat{\mathbf{y}} - \hat{\mathbf{x}}}{\sqrt{2}} \right) = E_0 \hat{\mathbf{b}}$, linearly polarized at −45º, as shown in Fig. 1b. As a consequence, when the nanoantenna is simultaneously fed by the two waveguides with controlled amplitudes and phases, a superposition of two orthogonal polarizations is radiated, as shown in Fig. 1c, with the electric field given by

$$\mathbf{E}_{rad} = E_0 (\alpha \hat{\mathbf{a}} + \beta \hat{\mathbf{b}}), \qquad (1)$$

where $\alpha$ and $\beta$ are the complex amplitudes feeding each input waveguide. The dual-input of this structure provides the two degrees of freedom required to cover the whole surface of the Poincaré sphere, spanning all possible polarization states of coherent plane waves.

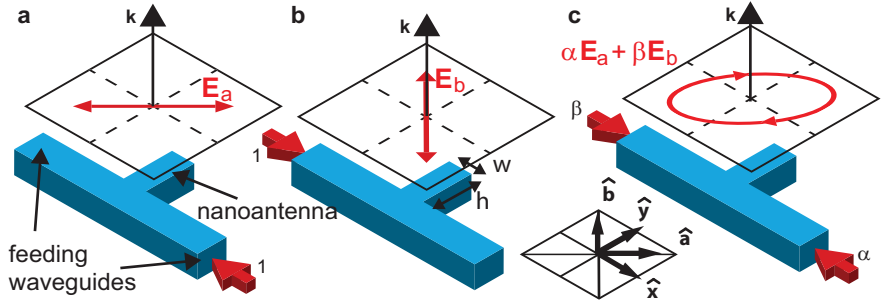

**Fig. 1. Polarization synthesis.** A dual-input nanoantenna radiating two orthogonal polarizations (**a**) and (**b**) when fed with each of its two inputs. The simultaneous feeding of both inputs results in a superposition of both cases (**c**). The dimensions of the fabricated and simulated nanoantennas are $w = 216$ nm and $h = 617$ nm.

To implement this method, we choose a silicon-on-insulator platform, which would allow the large-scale integration of nanoantennas using mainstream complementary metal oxide semiconductor (CMOS) technologies [14]. However, we stress that the method is universal and can be realized in any technological platform. In principle, any active nanophotonic system with tunable electro-optical modulation [23] would allow a straightforward electrically controlled phase and amplitude relation between both inputs, with a modulation bandwidth reaching tens of GHz on a silicon device [24], allowing ultrafast polarization tunability speed, fundamental for polarization scrambling or novel magnetic storage applications [8,9]. However, as a proof-of-principle we will demonstrate the effect in a simpler passive circuit by combining two effects: (i) the wavelength-dependent relative phase difference acquired by light propagating along waveguides with different path lengths (given by $\Delta\varphi = 2\pi n_{eff} \Delta L / \lambda_0$ where $\Delta L$ is the path length difference, $\lambda_0$ is the wavelength, and $n_{eff} = n_{eff}(\lambda_0)$ is the effective index of the mode) used to feed both inputs of the nanoantenna and (ii) we cascade broadband 50:50 Y-splitters to vary the relative amplitude between both inputs. In this way, with a given amplitude relation determined by the Y-splitters, we can vary the wavelength and thus sweep a circle in the surface of the Poincaré sphere of polarization states radiated in the normal direction by the nanoantenna, as shown in Fig. 2. For example, feeding both inputs simultaneously with an equal amplitude, and tuning the relative phase, we can sweep a full great circle of the Poincare sphere (Fig. 2b), including the synthesis of both linear polarizations ($E_x$ and $E_y$) and circular polarizations ($E_l$ and $E_r$), and a range of elliptical polarizations in between.

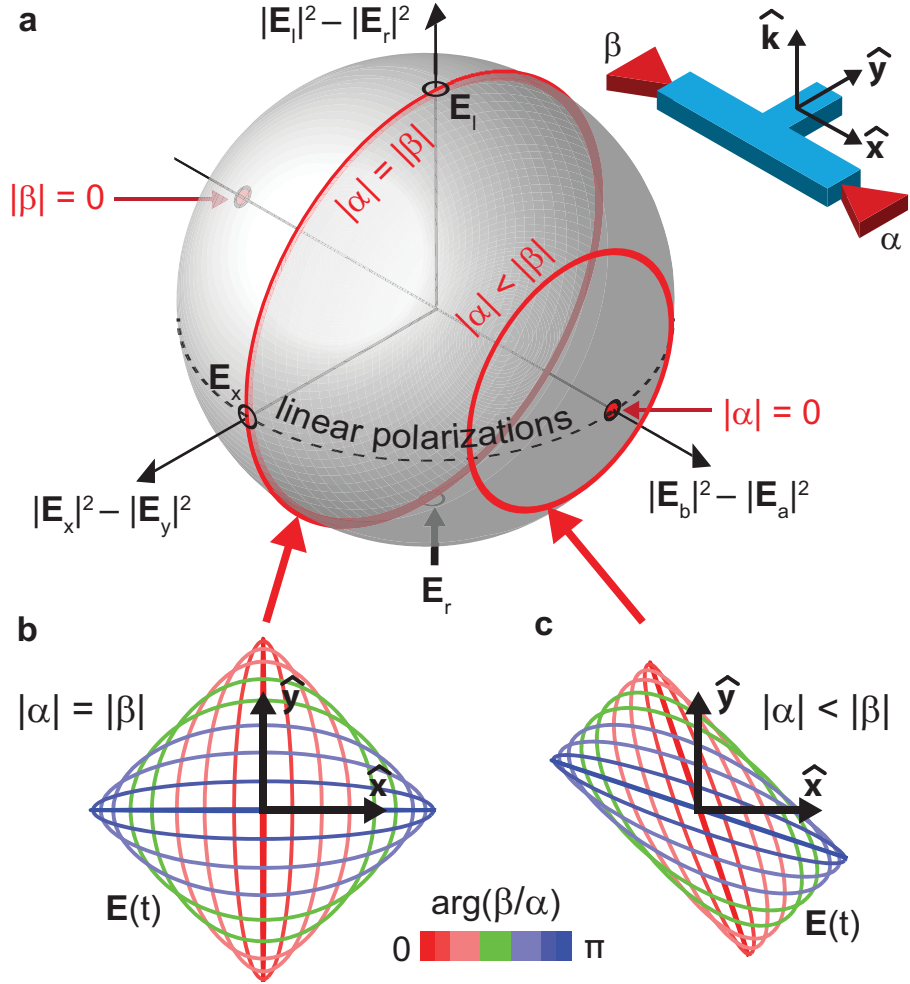

**Fig. 2. Poincaré sphere**. (**a**) The surface of the Poincaré sphere spans the whole space of possible polarizations of coherent plane waves, determined by three Stokes parameters $|E_a|^2 - |E_b|^2$, $|E_x|^2 - |E_y|^2$ and $|E_l|^2 - |E_r|^2$, normalized to the intensity $|E|^2$, where the subscripts refer to the components of the electric field in the bases being considered, where $\hat{l} = (\hat{x} + i\hat{y})/\sqrt{2}$ and $\hat{r} = (\hat{x} - i\hat{y})/\sqrt{2}$ correspond to the two circular polarization unit vectors. Our ideal nanoantenna radiates light polarized at +45º ($E_a$) or −45º ($E_b$) when fed by a single input. (**b**) Polarization ellipse for the simultaneous feeding of both inputs with an equal power (1:1) allowing the synthesis of a great circle of the Poincaré sphere by varying their relative phase. (**c**) Polarization ellipse for the simultaneous feeding of both inputs with a power ratio 4:1. The circle of attainable polarizations moves to the corresponding position on the axis $|E_b|^2-|E_a|^2$.

We experimentally tested this prediction using the setup shown in Fig. 3, where the dielectric nanoantenna is fed by two waveguides undergoing different propagation lengths after a symmetric Y-splitter ($|\alpha|=|\beta|$). The nanoantenna was designed as described in the supplementary information, and consists of a rectangular silicon protrusion of size $w$ = 216 nm, $h$ = 617 nm adjacent to a standard singlemode 250 x 400 nm silicon waveguide. A scanning electron microscope (SEM) image of the fabricated structure is included in Fig. 3. It has a broadband operation around 1550 nm, radiating light polarized at an angle of approximately +45º or –45º depending on the input being used (precisely, ±43.7º at 1540 nm, ±46.4º at 1550 nm, and ±49.8º at 1560 nm according to simulations). The percentage of power radiated in the undesired orthogonal polarization is less than 1% in the whole spectral range 1540 nm – 1560 nm. The effective area of the nanoantenna, defined as the power received by the nanoantenna divided by the power density incident on it from the normal direction, is 8390 nm$^2$ or 6% of the nanoantenna's physical area. Further optimization of this efficiency is desirable in practical applications, but it is more than enough for our proof-of-principle measurements. The radiated light in the normal direction is captured by a microscope after a linear polarizer, and the received intensity as a function of the polarizer angle and the wavelength is recorded, as shown in Fig. 4a, with the corresponding simulation shown in Fig. 4b. From these figures we can infer that the polarization state of radiated light changes from linear horizontal, to circular, to linear vertical, as the wavelength is changed. For added clarity, Fig. 4c maps the simulated radiated polarization at each wavelength into the surface of the Poincaré sphere. We see that the circle covered in the Poincaré sphere is very close to the great circle shown in Fig. 2a, but slightly tilted, as a result of minor variations of the fabricated device with respect to the ideal case whose radiated polarization is exactly ±45º at all wavelengths. In a device with an electrical control of the amplitude and phase of the optical signals reaching the nanoantenna, such variations could be easily corrected, which demonstrates the robustness of the method.

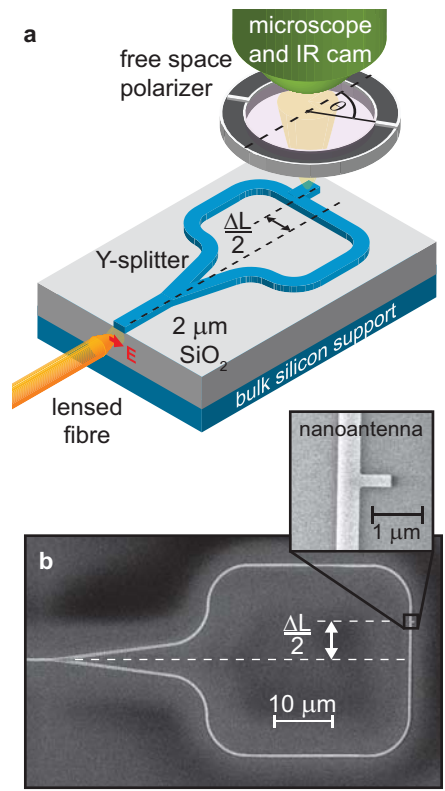

**Fig. 3. Experimental setup.** (**a**) Depiction of the experimental setup (not to scale) and (**b**) scanning electron microscope image of one of the fabricated structures. In the sample measured, $\Delta L \approx 20$ μm.

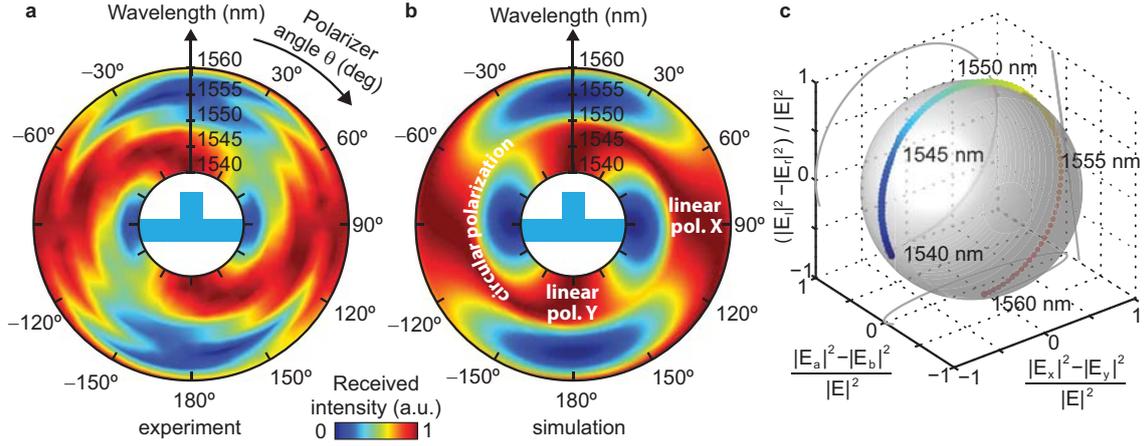

**Fig. 4. Experimental and simulation results.** (**a**) Received power at the detector as a function of laser wavelength and polarization filter rotation angle. (**b**) Corresponding electromagnetic simulation. (**c**) Mapping of the simulated polarizations into the Poincaré sphere. The simulations assumed a path length difference of $\Delta L = 19.88$ μm.

By cascading an additional Y-splitter on one of the feeding inputs (SEM image in Fig. 5a), the intensities $|\alpha|^2$ and $|\beta|^2$ follow a power ratio of 1:2, so the radiated polarization sweeps a different circle in the Poincaré sphere, as shown in Fig. 5b for the simulation of the actual nanoantenna. The experimental measurements are shown in Fig. 5c, and the corresponding simulation is given in Fig. 5d, showing a good correspondence, proving the synthesis of elliptically polarized light. Additional experimental results obtained from cascading three Y-splitters are presented in Fig. S1 of the supplementary materials. Breaking the nanoantenna symmetry along $\hat{\mathbf{y}}$ is a fundamental requirement for polarization synthesis, and a control experiment in which the nanoantenna is mirror symmetric in $\hat{\mathbf{y}}$ is also included in Fig. S1. Previous experiments showing unidirectional excitation of surface plasmons from a slit under circularly polarized illumination [19,20], although apparently symmetric in $\hat{\mathbf{y}}$, actually used grazing incidence as a fundamental requirement in order to break the symmetry.

In summary, polarization synthesis has been demonstrated here using nanoantennas in a CMOS compatible platform potentially allowing mass production. The general principle follows very simple symmetry considerations, and could be applied to more efficient and directive nanoantennas at any other frequency regime, material or technology, including plasmonics, fiber optics or even microwaves. This universal method is a very versatile and convenient way of achieving broadband polarization synthesis over the whole Poincaré sphere with a single nanoantenna element. The reciprocal setup, detecting different polarizations from the received amplitude and phase of the two output waveguides of a receiving nanoantenna, could be used for fast and integrated ellipsometers/polarimeters [25–27] and for the sorting of polarized photons impinging into a single nanoantenna. In general, as described in the supplementary information, the antenna could be designed for the ideal sorting of polarization states that are not necessarily orthogonal [28]. Together with the recent advances in the generation of complex beam shapes and polarizations, such as beams carrying orbital angular momentum [29,30], future integrated devices allowing an unprecedented control over polarization states of light are foreseen.

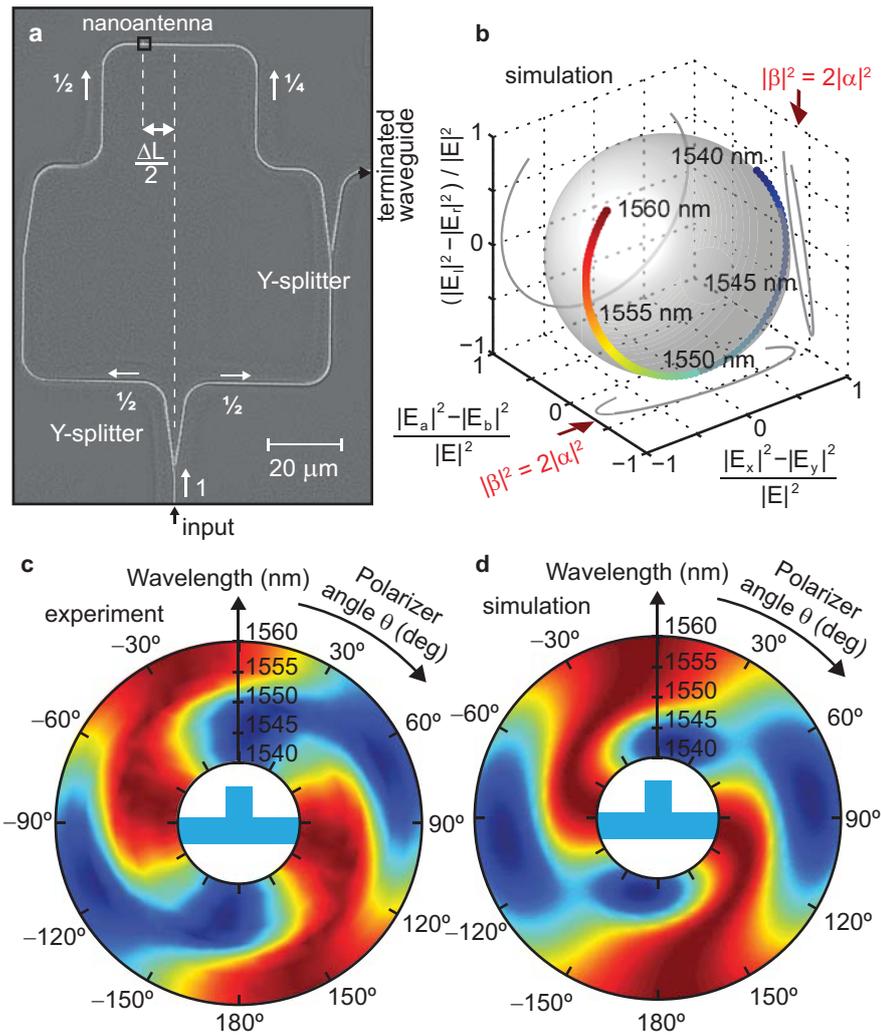

**Fig. 5. Experimental and simulation results using asymmetric excitation**. (**a**) Scanning electron microscope image of the fabricated structure used to achieve a 2:1 power ratio between both inputs. (**b**) Mapping of the resulting simulated polarizations into the Poincaré sphere. (**c**) Measured power at the detector as a function of laser wavelength and polarization filter rotation. (**d**) Corresponding electromagnetic simulation. The simulations assumed a path length difference of $\Delta L = 19.55$ μm.

**Acknowledgements:**

This work has received financial support from Spanish government (contracts Consolider EMET CSD2008-00066 and TEC2011-28664-C02-02). D. Puerto acknowledges support from grant Juan de la Cierva (JCI-2010-07479).

# Supplementary Materials for "Universal method for the synthesis of arbitrary polarization states radiated by a nanoantenna"

Francisco J. Rodríguez-Fortuño, Daniel Puerto, Amadeu Griol, Laurent Bellieres, Javier Martí and Alejandro Martínez

**Supplementary text:**

**Design of the nanoantenna**

The design of a nanoantenna which radiates, in the normal direction, a field polarized at plus or minus 45 degrees with respect to the symmetry plane, depending on the input waveguide being used, is straightforward if we consider the following symmetry arguments:

i. When the two feeding waveguides are excited with an electric field of equal amplitude and phase (an even symmetry in the electric field component $E_y$), then, due to the symmetry of the situation, the radiated light in the normal direction will be polarized only vertically $\mathbf{E_{rad}} = E_V \hat{\mathbf{y}}$, with a complex amplitude $E_V(\lambda)$.

ii. When the two feeding waveguides are excited with an electric field of equal amplitude but 180 degrees out of phase (odd symmetry in the electric field component $E_y$), then $E_y$ cancels out at the center, and the radiated light in the normal direction will be polarized only horizontally $\mathbf{E_{rad}} = E_H \hat{\mathbf{x}}$, with a complex amplitude $E_H(\lambda)$. This amplitude will be zero unless the mirror y-symmetry of the scatterer is broken.

The asymmetrical excitation of the scatterer by a single input waveguide can be seen as the linear superposition of the even (i) and odd (ii) scenarios described above, so the radiated light will be given by $\mathbf{E_{rad}} = \pm E_H \hat{\mathbf{x}} + E_V \hat{\mathbf{y}}$, where the $\pm$ sign depends on the waveguide being used. The quantities $E_H$ and $E_V$ can be easily obtained from fast numerical simulations exploiting the symmetry to reduce the numerical complexity. The optimization procedure then consists in making the two complex amplitudes equal $E_H(\lambda) = E_V(\lambda)$ in a broad frequency range.

**General case of an arbitrary nanoantenna**:

Here we will show the validity of our method in a more general case. We will only assume that the nanoantenna has two distinct input waveguides A and B which, when feeding the nanoantenna, radiate in a given direction **k** the polarized fields $\mathbf{E_A}$ and $\mathbf{E_B}$ respectively. No further assumptions in the symmetry or shape of the nanoantenna are required.

Our aim is to synthesize, in the radiated direction **k**, a desired polarization $\mathbf{E_{rad}}$, by using a linear combination of the feeding waveguides. The problem can be written as $\mathbf{E_{rad}} = \alpha \mathbf{E_A} + \beta \mathbf{E_B}$. In the far field, the radiated fields can be locally considered as plane waves with their electric field oscillating within the plane perpendicular to **k**. Therefore, the vectors $\mathbf{E_A}$, $\mathbf{E_B}$, and $\mathbf{E_{rad}}$ can be written in terms of a base of unit vectors $\hat{\mathbf{u}}$ and $\hat{\mathbf{v}}$ contained in such plane. This converts the problem into a linear system of two equations and two unknowns, namely

$$\begin{pmatrix} \mathbf{E_A} \cdot \hat{\mathbf{u}} & \mathbf{E_B} \cdot \hat{\mathbf{u}} \\ \mathbf{E_A} \cdot \hat{\mathbf{v}} & \mathbf{E_B} \cdot \hat{\mathbf{v}} \end{pmatrix} \cdot \begin{pmatrix} \alpha \\ \beta \end{pmatrix} = \begin{pmatrix} \mathbf{E_{rad}} \cdot \hat{\mathbf{u}} \\ \mathbf{E_{rad}} \cdot \hat{\mathbf{v}} \end{pmatrix}$$

which can be solved as long as the determinant of the matrix is nonzero, i.e. the vectors $\mathbf{E_A}$ and $\mathbf{E_B}$ are linearly independent, or, in physical terms, the fields $\mathbf{E_A}$ and $\mathbf{E_B}$ must not have the same polarization with only a phase and amplitude difference.

In the particular case presented in the manuscript, the unit vectors $\hat{\mathbf{u}}$ and $\hat{\mathbf{v}}$ can be identified with $\hat{\mathbf{x}}$ and $\hat{\mathbf{y}}$, and owing to the symmetry of the proposed structure, the system is given by:

$$\begin{pmatrix} E_H & -E_H \\ E_V & E_V \end{pmatrix} \cdot \begin{pmatrix} \alpha \\ \beta \end{pmatrix} = \begin{pmatrix} \mathbf{E}_x^{rad} \\ \mathbf{E}_y^{rad} \end{pmatrix}$$

Where the nanoantenna was designed so that $E_H = E_V = E_0/\sqrt{2}$, making the solution in that case simple and mathematically easy to describe.

**Supplementary figures:**

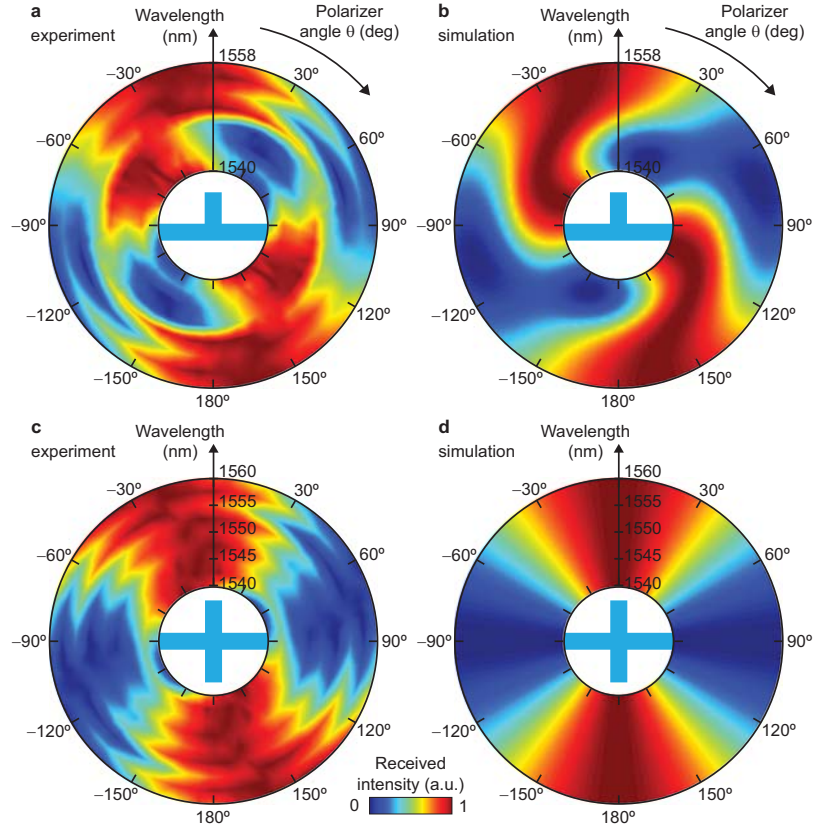

**Fig. S1. Experimental and simulation results using three cascaded Y-splitters on one feeding waveguide**. (**a**) Measured power at the detector as a function of laser wavelength and polarization filter rotation. (**b**) Corresponding electromagnetic simulation. (**c**) Measurement when the nanoantenna has mirror symmetry along **y**, radiating only vertical polarization regardless of the amplitude and phase of the inputs. (**d**) Corresponding electromagnetic simulation. The simulations assumed a path length difference of $\Delta L = 19.55$ μm.

**Materials and methods:**

**Simulations:** The optimization of the nanoantenna was performed using CST Microwave Studio and MATLAB. The simulations were performed with the nanoantenna as a receiver, with a normally incident plane wave of two orthogonal polarizations (x and y), and recording the complex spectrum of the mode amplitude excited in each of the two waveguides. The behavior as an emitting nanoantenna with any two feeding amplitudes is then immediately known by applying superposition and reciprocity principles. Simulations were performed in the time domain with a mesh step of λ/20, locally reduced to a very fine λ/150 mesh step within the nanoantenna and silicon waveguide.

**Fabrication:** The structures were fabricated on silicon-on-insulator (SOI) samples of SOITEC wafers with a top silicon layer thickness of 250 nm (resistivity ρ ~1-10 Ω cm$^{-1}$, with a lightly p-doping of ~10$^{15}$ cm$^{-3}$) and a buried oxide layer thickness of 3 μm. The structure fabrication is based on an electron beam direct writing process performed on a coated 100 nm hydrogen silsesquioxane resist film. This electron beam exposure, performed with a Raith150 tool, was optimized in order to reach the required dimensions employing an acceleration voltage of 30 KeV and an aperture size of 30 μm. After developing the HSQ resist using tetramethylammonium hydroxide as developer, the resist patterns were transferred into the SOI samples employing an also optimized Inductively Coupled Plasma- Reactive Ion Etching process with fluoride gases. Importantly, only a single lithography and etching step is needed to fabricate the structures, which stresses its simplicity.

**Measurements:** To obtain the output scattered power we used Xenics Xeva IR camera 1508 mounted on the eyepiece of a 4x microscope objective (National Stereoscopic Microscopes Zoom Models 420 Series) focused on the nanoantenna. The lensed fibre was aligned with the waveguide input to excite the TE mode, and the polarization of the scattered light was analysed with a free space polarizer before the microscope with the IR camera. We processed the images by integrating the camera counts in the scattering spot. To account for background noise, which imposes a background level of power, we subtract the background power per pixel (obtained from a nearby region without spot) to the spot power per pixel.